\documentclass[letterpaper, 10 pt, conference]{ieeeconf}  

\IEEEoverridecommandlockouts                              

\usepackage{graphics} 
\usepackage{epsfig} 
\usepackage{times} 
\usepackage{amsmath} 
\usepackage{amssymb}  
\usepackage{wrapfig}
\usepackage{sidecap}
\usepackage[font=footnotesize]{caption}

\title{\LARGE \bf Optimization theory of Hebbian/anti-Hebbian networks for PCA and whitening
}

\author{Cengiz Pehlevan$^{1}$ and Dmitri B. Chklovskii$^{1}$
\thanks{$^{1}$ Cengiz Pehlevan and Dmitri B. Chklovskii are with the Simons Center for Data Analysis, 160 Fifth Ave, New York, NY 10010. 
        {\tt\small {cpehlevan,mitya}@simonsfoundation.org}}%
}

\begin{document}

\maketitle
\thispagestyle{empty}
\pagestyle{empty}

\begin{abstract}

In analyzing information streamed by sensory organs, our brains face challenges similar to those solved in statistical signal processing. This suggests that biologically plausible implementations of online signal processing algorithms may model neural computation. Here, we focus on such workhorses of signal processing as Principal Component Analysis (PCA) and whitening which maximize information transmission in the presence of noise. We adopt the similarity matching framework, recently developed for principal subspace extraction, but modify the existing objective functions by adding a decorrelating term. From the modified objective functions, we derive online PCA and whitening algorithms which are implementable by neural networks with local learning rules, i.e. synaptic weight updates that depend on the activity of only pre- and postsynaptic neurons. Our theory offers a principled model of neural computations and makes testable predictions such as the dropout of underutilized neurons. 
\end{abstract}


\section{Introduction}

Principal Component Analysis (PCA) plays an important role in statistical signal processing by denoising data, identifying important features, and simplifying further processing. Mathematically, PCA computes the eigenvectors corresponding to the top eigenvalues of the data covariance matrix and projects data onto them. PCA algorithms exist for both the offline setting, where a whole dataset is available to the algorithm from the outset, and the online setting, where input data samples are presented to the algorithm sequentially, one at a time, and the corresponding output is computed prior to the presentation of the next input \cite{crammer2006online,arora2012stochastic,goes2014robust}. Given this, we expect that PCA algorithms, especially in the online setting, model some aspects of biological neural computation.

The first principal component of streamed data can be computed by a highly simplified model of a {\it single} neuron. Suppose that each input data sample is represented by the activity vector of upstream neurons at a corresponding time point. By summing these activities with the weights of corresponding synapses a neuron projects each data sample onto the vector of synaptic weights and transmits the projection to downstream neurons via its output activity. If synaptic weights are updated after each data sample presentation according to the Oja learning rule, a neuron computes the top eigenvector of the covariance matrix and outputs the first principle component \cite{oja1982simplified, hu2013neuron}. Here, we ignore temporal correlations in activity and assume that the dataset is presented as a sequence of static ``snapshots'' streamed in an arbitrary order.

The Oja learning rule has two major attractions for modeling neural computation. First, it can be derived, along with the weighted summation of inputs, from a principled objective function, by alternating minimization of a sum of squared representation errors with respect to activity and synaptic weights \cite{diamantaras1996principal}. Second, Oja learning is Hebbian, meaning that the weight update depends on the activity of only pre- and postsynaptic neurons, and hence biologically plausible. 

In order to extract {\it multiple} principal components from streamed data, researchers attempted to construct networks of multiple neurons, the activity of each representing a different principal component. However, most attempts have given up one of the two attractions of the single-neuron Oja rule. Instead of deriving a local learning rule from a principled objective function some researchers have simply postulated it \cite{sanger1989optimal,foldiak1989adaptive,rubner1989self,leen1990,kung1990neural}. Others, by minimizing the representation error, or its variants, derived single-layer neural networks with biologically implausible features such as nonlocal learning rules \cite{oja1992principal} or synapses that take part in  plasticity but not in neural dynamics \cite{leen1990}. 

Recently, we developed a novel theoretical framework, named similarity matching, that preserves both attractions of the single-neuron Oja rule in the multi-neuron case \cite{pehlevan2015MDS, pehlevan2014NMF, hu2014SMF, pehlevan2015NIPS}. Similarity matching postulates that similar inputs result in similar outputs and vice versa. Mathematically, pairwise similarities are quantified by the inner products of data vectors and matching is enforced by the classical multidimensional scaling (CMDS) cost function \cite{mardia1980multivariate}. We formulated a family of optimization problems and solved them in both offline and online settings. Importantly, the derived online algorithms correspond to a family of biologically plausible neural networks with local, Hebbian and anti-Hebbian, learning rules.

However, strictly speaking, the existing similarity matching algorithms \cite{pehlevan2015MDS, pehlevan2014NMF, hu2014SMF, pehlevan2015NIPS}, as well as many others \cite{foldiak1989adaptive,diamantaras1996principal}, do not perform PCA. Rather, they extract the principal subspace of the dataset, i.e. the space spanned by the eigenvectors corresponding to the top eigenvalues of the data covariance matrix, and project the data onto an arbitrary basis spanning this subspace (not necessarily the top eigenvectors {\it per se}). Yet, an algorithm to perform PCA is desirable because unlike other principal subspace projections its output is decorrelated.
 
Whitening, or equalization of variance across decorrelated output channels, is desirable because in the presence of Gaussian output noise and for limited output power, it achieves maximum information transmission \cite{linsker1988self,plumbley1993hebbian,plumbley1996information}. In neuroscience, the center-surround structure of retinal ganglion cell receptive fields is thought to implement whitening \cite{atick1992does,hyvarinen2009}.

In this paper, motivated by the optimal information transmission, we derive algorithms and networks for PCA and whitening in the similarity matching framework. The existing similarity matching objective functions \cite{pehlevan2015MDS,pehlevan2015NIPS} do not necessarily perform PCA because they depend only on the Grammian of the output and hence are invariant to orthogonal rotations of the output. As PCA is unique among principal subspace projections in that it produces a decorrelated output, we break the symmetry of the objective functions by adding a decorrelating term favoring PCA. 

We formulate and solve three optimization problems, each in online and offline settings. The solutions of the first and the second problems perform PCA of the input data. A common practice in PCA is to keep only a subset of principal components, containing the useful signal, for further processing. To this end, in the first problem, the number of output principle components is set by the smaller of the input and output numbers of channels. In the second problem, the number of output principle components is chosen adaptively, by hard-thresholding the eigenvalues of the data covariance matrix. The optimal solution of the third problem also chooses the number of output components adaptively by hard-thresholding but, in addition, whitens the output by equalizing the variance of orthogonal channels.

The paper is organized as follows. In Section \ref{sec2} we formulate optimization problems in the offline setting and present their solutions. In Section \ref{sec3} we derive corresponding online algorithms and demonstrate that they can be implemented by biologically plausible neural circuits. The performance of these online algorithms is evaluated numerically in Section \ref{sec4}. In Section \ref{sec5} we predict that underutilized neurons drop out of the circuit. Section \ref{sec6} comments on decorrelating interneuron activities.

\begin{SCfigure*}
\centering
\includegraphics[scale=0.85]{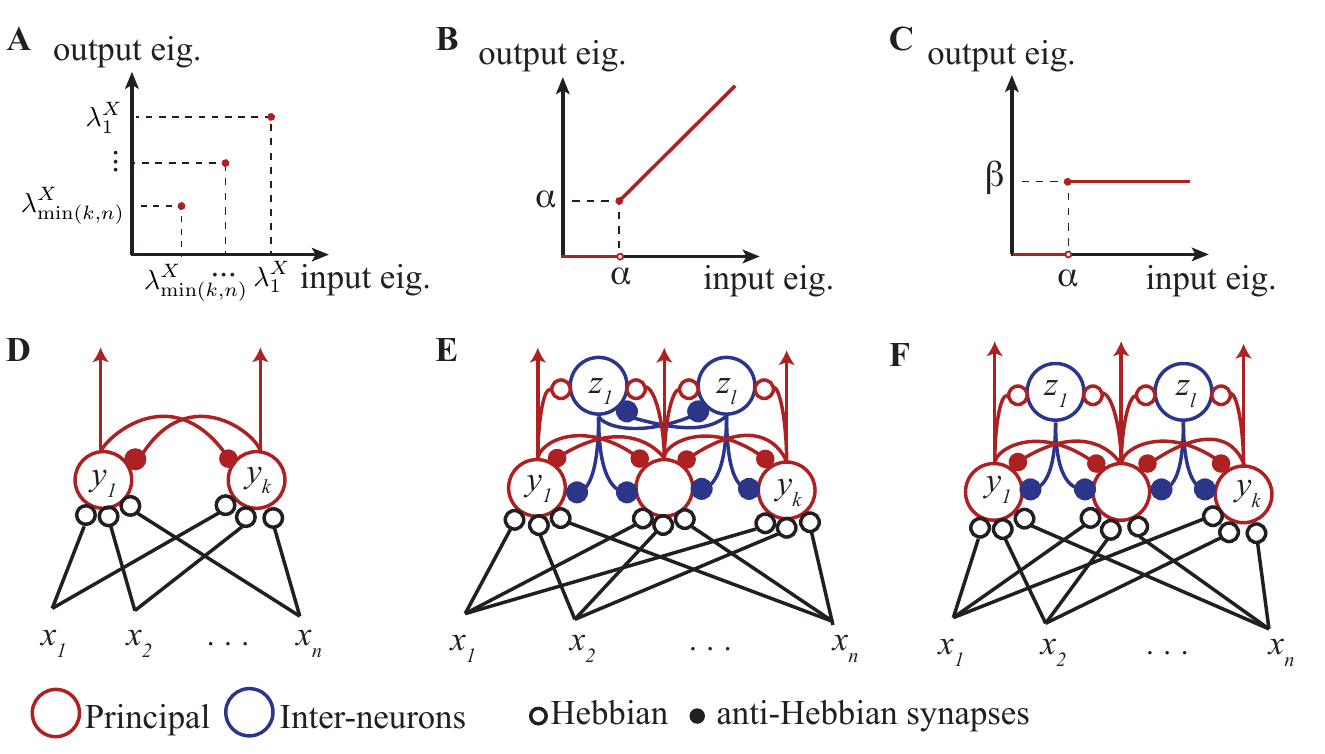}
\captionsetup{labelformat=empty}
 \caption{\footnotesize Fig. 1: Input-output functions of the three offline solutions and neural network implementations of the corresponding online algorithms. {\bf A-C.} Input-output functions of covariance eigenvalues. {\bf A.} Thresholding of top $\min(k,n)$ components. {\bf B.} Adaptive hard-thresholding. {\bf C.} Whitening after thresholding. {\bf D-F.} Corresponding network architectures. \label{Fig1}}
\end{SCfigure*}

\section{PCA and whitening in the offline setting}\label{sec2}
In this Section, we introduce and solve three novel optimization problems in the offline setting:
\begin{align}
{\rm Offline \: setting:} \; {\bf Y} \leftarrow  \mathop {\arg \min }\limits_{{\bf Y}}\,L\left({\bf X},{\bf Y}\right),
\end{align}
where the input, ${\bf X}=[{\bf x}_1,\ldots,{\bf x}_T]$ is an $n\times T$ matrix with $T$ centered input data samples in $\mathbb{R}^n$ as its columns and the output, ${\bf Y}=[{\bf y}_1,\ldots,{\bf y}_T]$ is a $k\times T$  matrix with corresponding outputs in $\mathbb{R}^k$ as its columns. In this Section, we assume $T\geq n$ and $T\geq k$ for convenience, however our results could be generalized easily.

\subsection{Similarity matching cost function for PCA}
To formulate similarity matching mathematically, we minimize the summed squared differences between all pairwise similarities, the so-called CMDS cost function \cite{mardia1980multivariate,pehlevan2015NIPS}:
\begin{align}\label{ST}
\min_{\mathbf{Y}} \left\| \mathbf{X}^\top \mathbf{X} - \mathbf{Y}^\top\mathbf{Y} \right\|_F^2.
\end{align}

Optimal solutions of the CMDS cost function \eqref{ST} are  projections of the input dataset $\mathbf{X}$ onto its principal subspace \cite{mardia1980multivariate,pehlevan2015MDS}. Suppose the eigen-decomposition of  $\mathbf{X}^\top\mathbf{X} = \mathbf{V}^X \mathbf{\Lambda}^X {\mathbf{V}^X}^\top$, where $\mathbf{\Lambda}^X = \text{diag}(\lambda_1^X, ..., \lambda_T^X)$ with $\lambda_1^X \geq ... \geq \lambda_T^X \geq 0$ are ordered eigenvalues of $\mathbf{X}^\top{\bf X}$.  Then the following is a solution of \eqref{ST}: 
\begin{align}\label{STsol}
\mathbf{Y} = \mathbf{U}_k{ \mathbf{\Lambda}^Y_k}^{1/2} {\mathbf{V}_k^X}^\top,
\end{align}
where $\mathbf{\Lambda}^Y_k$ is a $k \times k$ diagonal matrix whose non-zero diagonals are $\lbrace\lambda_1^X, ..., \lambda_{\min(k,n)}^X\rbrace$, ${\bf V}_k^X$ consists of the columns of ${\bf V}^X$ corresponding to the top $k$ eigenvalues, ${\bf V}^X_k = \left[{\bf v}^X_{1},\ldots,{\bf v}^X_{k}\right]$, and ${\bf U}_k$ is any $k \times k$ orthogonal matrix.

However, the solution of \eqref{ST} is not unique: because the objective function depends on the output only via its Grammian, it is invariant to an orthogonal left-rotation of $\mathbf{Y}$. This degree of freedom corresponds to an arbitrary choice of an orthogonal matrix ${\bf U}_k$ in \eqref{STsol}. 

To eliminate this degree of freedom we take advantage of the fact that the only way to decorrelate the output,  i.e. obtain diagonal output covariance matrix, is to compute {\it bona fide} principal components. Thus, we add to the objective the sum squared of the off-diagonal elements of the output covariance matrix, $\frac 1T{\bf Y} {\bf Y}^\top = \frac 1T {\bf U}_k{\bf \Lambda}^Y_k{\bf U}_k^\top$: 
\begin{align}\label{STreg}
 \min_{\bf Y} & \left [ \left\Vert {\bf X}^\top{\bf X}-{\bf Y}^\top{\bf Y}\right\Vert_F^2  +\gamma\left\Vert {\rm off}\left({\bf Y} {\bf Y}^\top\right) \right\Vert_F^2\right],
 \end{align}
where $\gamma>0$ and the operator ${\rm off}()$ extracts off-diagonal elements of a matrix by setting diagonal elements to $0$.

Eq. \eqref{STreg} defines an objective function for PCA, where data is projected onto top $\min(k,n)$ principal eigenvectors. Indeed, for correlated solutions \eqref{STsol} $\gamma \left\Vert {\rm off}\left({\bf Y} {\bf Y}^\top\right) \right\Vert_F^2$ is positive, thus resulting in suboptimal values of the objective. For uncorrelated solutions, $\gamma \left\Vert {\rm off}\left({\bf Y} {\bf Y}^\top\right) \right\Vert_F^2$ vanishes and therefore does not affect the value of the objective.  Note that, if $k > n$ decorrelation implies that $k-n$ output channels are silent. 

\subsection{Objective function for adaptive PCA}

One drawback of the PCA objective function \eqref{STreg} is that the number of output dimensions must be chosen prior to the presentation of the first data sample. In real-life situations, and, especially, neuroscience context, the input signal-to-noise ratio may not be known and the output dimensionality must adapt automatically. Such adaptive dimensionality reduction solution may be obtained by solving the following minimax problem \cite{pehlevan2015NIPS}:
\begin{align}\label{HT}
 \min_{{\bf Y}}  \max_{{\bf Z}} &\left[\left\Vert {\bf X}^\top{\bf X}-{\bf Y}^\top{\bf Y}\right\Vert_F^2  -  \left\Vert {\bf Y}^\top{\bf Y}-{\bf Z}^\top{\bf Z}\right\Vert_F^2\right. \nonumber \\ &\qquad\left.+ 2\alpha T\, {\rm Tr}\left({\bf Y}^\top{\bf Y}\right)- 2\alpha T \,{\rm Tr}\left({\bf Z}^\top{\bf Z}\right)\right], 
 \end{align}
where we introduced an auxiliary variable ${\bf Z}=[{\bf z}_1,\ldots,{\bf z}_T]\in \mathbb{R}^{lxT}$. 

The number of output dimensions, $m={\rm rank}({\bf Y})$, is determined by the trade off between the similarity matching and the regularization terms. Whereas higher rank reduces the matching error it adds to the regularizer, Tr$\left(\mathbf{Y}^\top\mathbf{Y}\right)$, because the regularizer is a nuclear norm of the Grammian, $\mathbf{Y}^\top\mathbf{Y}$, which is a convex relaxation of rank. Then, the number of output dimensions, $m$, is the number of eigenvalues of the input data covariance matrix, ${\bf C} = \frac 1T {\bf X}{\bf X}^\top$, greater than or equal to $\alpha>0$. We assume that $k\geq m$ and $l \geq m$.

Objective \eqref{HT} is solved by projecting the input dataset onto the $m$-dimensional principal subspace of input covariance \cite{pehlevan2015NIPS}. Specifically, suppose the eigen-decomposition of ${\bf X}^\top{\bf X} = {\bf V}^X {\bf \Lambda}^X {{\bf V}^X}^\top$ where $\mathbf{\Lambda}^X = \text{diag}(\lambda_1^X, ..., \lambda_T^X)$ with $\lambda_1^X \geq ... \geq \lambda_T^X \geq 0$ are ordered eigenvalues of $\mathbf{X}^\top {\bf X}$, as before. Then, the optimal ${\bf Y}$ and ${\bf Z}$ are:
 \begin{align}\label{hYZ}
{\bf Y} &= {\bf U}_k\,{\bf HT}_k({\bf \Lambda}^X,\alpha T)^{1/2} \, {{\bf V}_k^X}^\top,\nonumber \\ {\bf Z} &= {\bf U}_l\,{\bf ST}_l({\bf \Lambda}^X,\alpha T)^{1/2} \, {{\bf V}_l^X}^\top,
\end{align}
where ${\bf HT}_k({\bf \Lambda}^X,\alpha T)  = {\rm diag} \left({\rm HT}\left(\lambda^X_{1},\alpha T\right),\ldots,{\rm HT}\left(\lambda^X_{k},\alpha T \right)\right)$, {\rm HT} is the hard-thresholding function, ${\rm HT}(a,b) = a\Theta(a-b)$ with $\Theta()$ being the step function: $\Theta(a-b)=1$ if $ a \ge b$ and $\Theta(a-b)=0$ if $a<b$, ${\bf ST}_l({\bf \Lambda}^X,\alpha T)  = {\rm diag} \left({\rm ST}\left(\lambda^X_{1},\alpha T\right),\ldots,{\rm ST}\left(\lambda^X_{l},\alpha T \right)\right)$, {\rm ST} is the soft-thresholding function, ${\rm ST}(a,b) = \max(a-b,0)$, ${\bf V}^X_p = \left[{\bf v}^X_{1},\ldots,{\bf v}^X_{p}\right]$ and ${\bf U}_p$ is any $p \times p$ orthogonal matrix. 

Similarly to the observation in the previous subsection, the solution of \eqref{HT} is not unique because the objective function is invariant to orthogonal left-rotations of $\mathbf{Y}$. To obtain PCA as the unique optimal solution, as before, we add a term to the objective function that penalizes off-diagonal elements of the covariance matrix: 
\begin{align}\label{HTreg}
& \min_{{\bf Y}}  \max_{{\bf Z}} \left[\left\Vert {\bf X}^\top{\bf X}-{\bf Y}^\top{\bf Y}\right\Vert_F^2  -  \left\Vert {\bf Y}^\top{\bf Y}-{\bf Z}^\top{\bf Z}\right\Vert_F^2\right. \nonumber \\ &\left.+ 2\alpha T\, {\rm Tr}\left({\bf Y}^\top{\bf Y}\right)- 2\alpha T \,{\rm Tr}\left({\bf Z}^\top{\bf Z}\right) + \gamma \left\Vert {\rm off}\left({\bf Y} {\bf Y}^\top\right) \right\Vert_F^2\right], 
 \end{align}
where $\gamma>0$. 

Eq. \eqref{STreg} defines an objective function for PCA, where data is projected onto top $m$ principal eigenvectors. Indeed, for correlated solutions \eqref{hYZ},  $\gamma \left\Vert {\rm off}\left({\bf Y} {\bf Y}^\top\right) \right\Vert_F^2$ is positive, thus resulting in suboptimal values of the objective. For uncorrelated solutions, $\gamma \left\Vert {\rm off}\left({\bf Y} {\bf Y}^\top\right) \right\Vert_F^2$ vanishes and does not affect the value of the objective. If the number of eigenvalues of $\frac 1T{\bf X}{\bf X}^\top$ greater than or equal to $\alpha$ is less than the number of output channels, $m<k$, decorrelation implies that some output channels will be silent.

\subsection{Objective function for whitening}

Next we consider an objective function that leads to equalization of the ouptut covariance eigenvalues after thresholding \cite{pehlevan2015NIPS}:
\begin{align}\label{whiten}
 \min_{{\bf Y}}  \max_{{\bf Z}} \,& {\rm Tr}\left(- {\bf X}^\top{\bf X}{\bf Y}^\top{\bf Y}   +\alpha T{\bf Y}^\top{\bf Y} \right. \nonumber \\
&\qquad\qquad \left.+ {\bf Y}^\top{\bf Y}{\bf Z}^\top{\bf Z}  - \beta T{\bf Z}^\top{\bf Z}\right),
 \end{align}
where $\alpha>0$ controls the number of degrees of freedom in the output, $m$, and $\beta >0$ sets magnitude of output eigenvalues. As before, we assume that $k\geq m$ and $l\geq m$.
 
Objective \eqref{whiten} is optimized by projecting the input dataset onto its principal subspace and equalizing the non-zero eigenvalues \cite{pehlevan2015NIPS}. Specifically, suppose the eigen-decomposition of ${\bf X}^\top{\bf X} = {\bf V}^X {\bf \Lambda}^X {{\bf V}^X}^\top$, where ${\bf \Lambda}^X={\rm diag}\left(\lambda^X_{1},\ldots,\lambda^{X}_T\right)$ with $\lambda^{X}_1\geq\ldots\geq \lambda^{X}_T\geq 0$.  Then, the optimal ${\bf Y}$ and ${\bf Z}$ are:
 \begin{align}\label{wYZ}
{\bf Y} &={\bf U}_k \,\sqrt{\beta T}\, {\bf \Theta}_k({\bf \Lambda}^X,\alpha T) \, {{\bf V}^X_k}^\top,\nonumber \\ {\bf Z} &= {\bf U}_l \, {\bf \Sigma}_l \, {\bf \Theta}_l({\bf \Lambda}^X,\alpha T) \, {{\bf V}^X_l}^\top,
\end{align}
where  ${\bf \Sigma}_l = \,{\rm diag} \left(\sigma_1,\ldots,\sigma_l\right)$ with $\sigma_i$ arbitrary constants, ${\bf \Theta}_k({\bf \Lambda}^X,\alpha T) =  {\rm diag} \left(\Theta\left(\lambda^X_{1}-\alpha T\right),\ldots,\Theta\left(\lambda^X_{k}-\alpha T \right)\right)$, ${\bf V}_p = \left[{\bf v}^X_1,\ldots,{\bf v}^X_p\right]$ and  ${\bf U}_p$  is any $p \times p$ orthogonal matrix. There are other optimal ${\bf Z}$, see \cite{pehlevan2015NIPS} for full expressions.

As before, the solution of \eqref{whiten} is not unique. Even though eigenvalues are equalized, due to the freedom in choosing ${\bf U}_k$, the variances of output channels are not equal, generally. An exception is the case where $k=l=m$. Then ${\bf Y}$ is full-rank and $\frac{1}T {\bf Y}{\bf Y}^\top = \beta{\bf I}_k$, implying that the output is whitened.  To obtain whitening as the unique optimal solution for general $k\geq m$ and $l\geq m$, following the arguments of the previous subsections, we add a term to the objective function that penalizes off-diagonal elements of the covariance matrix: 
\begin{align}\label{whitenreg}
 \min_{{\bf Y}}  \max_{{\bf Z}} \, &\left[{\rm Tr}\left(- {\bf X}^\top{\bf X}{\bf Y}^\top{\bf Y} +\alpha T{\bf Y}^\top{\bf Y}+ {\bf Y}^\top{\bf Y}{\bf Z}^\top{\bf Z} \right.\right. \nonumber \\
&\left. \left.  \qquad\quad- \beta T{\bf Z}^\top{\bf Z} \right) + \frac \gamma 2 \left\Vert {\rm off}\left({\bf Y} {\bf Y}^\top\right) \right\Vert_F^2 \right],
 \end{align}
where $\gamma>0$. 

Eq. \eqref{whitenreg} defines an objective function for whitening, which can be solved by projecting the input dataset onto top $m$ principal eigenvectors with variance in each channel normalized to $\beta$. Indeed, for correlated solutions \eqref{wYZ}, $(\gamma/2)\left\Vert {\rm off}\left({\bf Y} {\bf Y}^\top\right) \right\Vert_F^2$ is positive, thus resulting in suboptimal values of the objective. For uncorrelated solutions, $(\gamma/2) \left\Vert {\rm off}\left({\bf Y} {\bf Y}^\top\right) \right\Vert_F^2$ vanishes and does not affect the value of the objective.  Note that, if the number of eigenvalues of $\frac 1T{\bf X}{\bf X}^\top$ greater than or equal to $\alpha$, is less than the number of output channels, $m<k$, decorrelation implies that some output channels will be silent.

\section{Online learning rules for decorrelated output}\label{sec3}

Unlike the offline setting where the whole input dataset is available before an output is computed, neurons compute output, ${\bf y}_T$, for each data sample presentation, ${\bf x}_T$, before the next data sample is presented and past outputs cannot be altered.  Therefore, we formulate optimization problems in the online setting where optimization must be performed at every time step, $T$, on the objective which is a function of inputs and outputs up to time, $T$:
\begin{align}
 {\rm Online \: setting:} \; {\bf y}_T \leftarrow  \mathop {\arg \min }\limits_{{\bf y}_T}\,L\left({\bf X},{\bf Y}\right).
\end{align}

In this Section, we solve the three optimization problems in the online setting and map the steps of the online algorithms onto the dynamics of neuronal activity and local learning rules for synaptic weights. Our derivations follow the methods described in detail \cite{pehlevan2015MDS,pehlevan2015NIPS}.

\subsection{Online similarity matching for PCA}

We start with an online version of the objective function \eqref{STreg}:
\begin{align}\label{STstrainOnline}
{\bf y}_T \leftarrow   \mathop {\arg \min }\limits_{{\bf y}_T} \left[ \left\Vert {\bf X}^\top{\bf X}-{\bf Y}^\top{\bf Y}\right\Vert_F^2  + \gamma \left\Vert {\rm off}\left({\bf Y} {\bf Y}^\top\right) \right\Vert_F^2\right]. 
 \end{align}
By expanding the squared Frobenius norms and keeping only the terms that depend ${\bf y}_T$ we get:
\begin{align}\label{onlineStrain_old} 
{\bf y}_T \leftarrow &\mathop {\arg \min }\limits_{{\bf y}_T} \left[  - 4{{\bf x}^\top_T}\left( {\sum\limits_{t = 1}^{T - 1} {{\bf x}_t{{\bf y}^\top_t}} } \right){\bf y}_T \right. \nonumber \\ &\left. + 2{{\bf y}^\top_T}\left( {\sum\limits_{t = 1}^{T - 1} {{\bf y}_t{{\bf y}^\top_t}}+ }  \gamma \, {\rm off}\left({\sum\limits_{t = 1}^{T - 1} {{\bf y}_t{{\bf y}^\top_t}} }\right) \right){\bf y}_T \right.\nonumber\\
& \left. - 2{{\left\| {{\bf x}_T} \right\|}^2}{{\left\| {{\bf y}_T} \right\|}^2} + {\left\| {{\bf y}_T} \right\|}^4 + \gamma {{\bf y}^\top_T}\, {\rm off}\left({{{\bf y}_T{{\bf y}^\top_T}} }\right) {\bf y}_T\right].
\end{align}
In the large-$T$ limit, the first two terms grow linearly with $T$ and dominate over the last three terms which can be dropped. The remaining objective is a positive definite quadratic form of ${\bf y}_T$ and the optimization problem is convex. At its minimum, the following holds:
\begin{align}
\left( {\sum\limits_{t = 1}^{T - 1} {{\bf y}_t{{\bf y}^\top_t}} } + \gamma \, {\rm off}\left({\sum\limits_{t = 1}^{T - 1} {{\bf y}_t{{\bf y}^\top_t}} }\right)\right){\bf y}_T = \left( {\sum\limits_{t = 1}^{T - 1} {{\bf y}_t{{\bf x}^\top_t}} } \right) {\bf x}_T.
\end{align}

We could solve for ${\bf y}_T$ analytically via matrix inversion, however, to obtain a neurally plausible algorithm, we solve these equations by a weighted Jacobi iteration\footnote{See \cite{pehlevan2015MDS} for other possible iterative solutions and their convergence properties}:
\begin{align}\label{dynamics}
{\bf y}_{T} \leftarrow  \left(1-\eta \right){\bf y}_{T} + \eta \left({\bf W}^{YX}_{T}{\bf x}_{T} - {\bf W}^{YY}_{T} {\bf y}_{T}\right).
\end{align}
where $\eta$ is the weight parameter, and ${\bf W}^{YX}_T$ and ${\bf W}^{YY}_T$ are normalized input-output and output-output covariances, 
\begin{align}\label{WM}
W^{YX}_{T,ij} &= {{\sum\limits_{t = 1}^{T - 1} {y_{t,i}^{}x_{t,j}^{}} }}\bigg/{\sum\limits_{t = 1}^{T - 1} {y_{t,i}^2} },\nonumber \\  W^{YY}_{T,i,j \ne i} &= {(1+\gamma)\sum\limits_{t = 1}^{T - 1} y_{t,i}y_{t,j} }\bigg/{ \sum\limits_{t = 1}^{T - 1} {y_{t,i}^2} },\quad W^{YY}_{T,ii} = 0.
\end{align}

Remarkably, iteration \eqref{dynamics} can be implemented by neuronal dynamics in a single-layer network, Figure 1D. In this interpretation,  ${\bf W}^{YX}_T$ and ${\bf W}^{YY}_T$ represent the weights of feedforward (${\bf x}_t\rightarrow{\bf y}_t$) and lateral (${\bf y}_t\rightarrow{\bf y}_t$) synaptic connections, respectively. Interestingly, although the optimization problems \eqref{ST} and \eqref{STstrainOnline} are formulated only in terms of input and output activities, we recovered expressions naturally identified as feedforward and lateral synaptic weights. 

At each data sample presentation, $T$, after the output ${\bf y}_T$ converges to a steady state, synaptic weights are updated according to \eqref{WM}. By rewriting \eqref{WM} in a recursive form, we can eliminate the need to keep all past input and output in memory  and obtain a fully online algorithm. To this end, let us define a scalar variable ${D}^Y_{T,i}$ representing cumulative activity of a neuron $i$ up to time $T-1$, 
\begin{align}\label{D}
{D}^Y_{T,i} =\sum\limits_{t = 1}^{T - 1} {y_{t,i}^2}.
\end{align}
Then, synaptic weight updates are:
\begin{align}\label{hah}
D^Y_{T+1,i} &\leftarrow D^Y_{T,i}+ y_{T,i}^2\nonumber\\
{W^{YX}_{T+1,ij}} &\leftarrow {W^{YX}_{T ,ij}} + \left( y_{T,i}x_{T,j} - y_{T,i}^2{W^{YX}_{T,ij}} \right)/D^Y_{T+1,i} \nonumber\\
{W^{YY}_{T+1,i,j \ne i}} &\leftarrow {W^{YY}_{T,ij}} \nonumber \\ & \qquad+  \left( (1+\gamma)y_{T,i}y_{T,j} -  y_{T,i}^2 {W^{YY}_{T,ij}} \right)/D^Y_{T+1,i} .
\end{align}

To summarize, equations \eqref{dynamics} and \eqref{hah} define a neural network algorithm that solves the optimization problem \eqref{STstrainOnline} for streaming data by alternating between two phases: neural activity dynamics and synaptic updates. After a data sample is presented at time $T$, the algorithm goes into the neuronal activity phase \eqref{dynamics}, where neuron activities are updated until convergence to a fixed point. Then, in the second phase of the algorithm, synaptic weights are updated, according to a local Hebbian rule \eqref{hah} for feedforward connections, and according to a local anti-Hebbian rule (due to the $(-)$ sign in equation \eqref{dynamics}) for lateral connections. Interestingly, these updates have the same form as the single-neuron Oja's rule \cite{oja1982simplified}, except that the learning rate is not a free parameter but is determined by the cumulative neuronal activity $1/D^Y_{T+1,i}$. 

A similar network was derived in \cite{pehlevan2015MDS} from the objective \eqref{STstrainOnline} without the decorrelating term, i.e. $\gamma = 0$. The addition of the decorrelating term did not spoil the locality of learning rules, nor did it change the network architecture. The only difference is the strengthening of lateral synaptic weights by a factor $(1+\gamma)$ \eqref{WM}. Lateral connections implement competition between the output of neurons: without them, $k$ output neurons would independently recover the first principal component \cite{oja1982simplified}. Interestingly, the strengthening of lateral synapses is sufficient to decorrelate neuronal output and project the input to its principal eigenvectors, as opposed to an arbitrary basis in the principal subspace, as was the case in \cite{pehlevan2015MDS}.

\subsection{Online adaptive PCA}

Next, we consider an online version of \eqref{HT}:
\begin{align}\label{HTOnline}
&\{{\bf y}_T,{\bf z}_T\} \leftarrow \mathop {\arg \min }\limits_{{\bf y}_T}  \mathop {\arg \max }\limits_{{\bf z}_T}  \left[\left\Vert {\bf X}^\top{\bf X}-{\bf Y}^\top{\bf Y}\right\Vert_F^2\right.  \nonumber \\
& \left.-  \left\Vert {\bf Y}^\top{\bf Y}-{\bf Z}^\top{\bf Z}-\alpha T{\bf I}_T \right\Vert_F^2 + \gamma\left\Vert {\rm off}\left({\bf Y} {\bf Y}^\top\right) \right\Vert_F^2\right]. 
 \end{align}
By expanding the norms and keeping only those terms that depend on ${\bf y}_T$ or ${\bf z}_T$ and taking the large-$T$ limit, we get:
\begin{align}\label{HTOnlineReduced}
&\{{\bf y}_T,{\bf z}_T\}\leftarrow  \mathop {\arg \min }\limits_{{\bf y}_T}  \mathop {\arg \max }\limits_{{\bf z}_T} \left[ - 4{{\bf x}^\top_T}\left( {\sum\limits_{t = 1}^{T - 1} {{\bf x}_t{{\bf y}^\top_t}} } \right){\bf y}_T \right.\nonumber \\
&\left. +2{{\bf y}^\top_T}\left( \gamma\,{\rm off}\left({\sum\limits_{t = 1}^{T - 1} {{\bf y}_t{{\bf y}^\top_t}} }\right) +\alpha T{\bf I}_k \right){\bf y}_T     \right. \nonumber \\
&\left.+ 4{{\bf y}^\top_T}\left( {\sum\limits_{t = 1}^{T - 1} {{\bf y}_t{{\bf z}^\top_t}} } \right){\bf z}_T - 2{{\bf z}^\top_T}\left( {\sum\limits_{t = 1}^{T - 1} {{\bf z}_t{{\bf z}^\top_t}} }+\alpha T{\bf I}_l \right){\bf z}_T \right].
 \end{align}
This objective is a convex quadratic from in ${\bf y}_T$ and concave quadratic form in ${\bf z}_T$. The solution of this minimax problem is a saddle-point of the objective function, which is found by setting the gradient of the objective with respect to $\lbrace{\bf y}_T,{\bf z}_T\rbrace$ to zero \cite{boyd2004convex}:
\begin{align}
\left( \gamma\,{\rm off} \left({\sum\limits_{t = 1}^{T - 1} {{\bf y}_t{{\bf y}^\top_t}} }\right) +\alpha T {\bf I}_k \right){\bf y}_T &= \left( {\sum\limits_{t = 1}^{T - 1} {{\bf y}_t{{\bf x}^\top_t}} } \right){\bf x}_T \nonumber \\ &\quad - \left( {\sum\limits_{t = 1}^{T - 1} {{\bf y}_t{{\bf z}^\top_t}} } \right){\bf z}_T, \nonumber \\
\left( {\sum\limits_{t = 1}^{T - 1} {{\bf z}_t{{\bf z}^\top_t}} }+\alpha T {\bf I}_l  \right){\bf z}_T& = \left( {\sum\limits_{t = 1}^{T - 1} {{\bf z}_t{{\bf y}^\top_t}} } \right){\bf y}_T.
\end{align}
We could solve these linear equations analytically, but to obtain a neurally plausible algorithm, we solve them using a weighted Jacobi iteration:
\begin{align}\label{HTdynamics}
{\bf y}_{T} &\leftarrow  \left(1-\eta \right){\bf y}_{T} + \eta \left({\bf W}^{YX}_{T}{\bf x}_{T}  - {\bf W}^{YZ}_{T} {\bf z}_{T}- {\bf W}^{YY}_{T} {\bf y}_{T}\right), \nonumber \\
 {\bf z}_{T} &\leftarrow  \left(1-\eta \right){\bf z}_{T} + \eta \left({\bf W}^{ZY}_{T}{\bf y}_{T} - {\bf W}^{ZZ}_{T} {\bf z}_{T}\right), 
\end{align}
where $\eta$ is the weight parameter and
\begin{align}\label{HTWM}
W^{YX}_{T,ij} &= \frac1 {\alpha T}  {{\sum\limits_{t = 1}^{T - 1} {y_{t,i}^{}x_{t,j}^{}} }},  \quad {W}^{YZ}_{T,ij} = \frac1 {\alpha T}{{\sum\limits_{t = 1}^{T - 1} {y_{t,i}^{}z_{t,j}^{}} }} \nonumber \\ W^{YY}_{T,i,j\neq i} &= \frac {\gamma} {\alpha T}{ {\sum\limits_{t = 1}^{T - 1} {y_{t,i}y_{t,j}} }}, \qquad {W}^{YY}_{T,ii} = 0,\nonumber \\
W^{ZY}_{T,ij} &= {{\sum\limits_{t = 1}^{T - 1} z_{t,i}y_{t,j} }}\bigg/{\left(\alpha T + \sum\limits_{t = 1}^{T - 1} {z_{t,i}^2} \right)}, \nonumber \\ W^{ZZ}_{T,i,j \neq i} &= {{\sum\limits_{t = 1}^{T - 1} z_{t,i}z_{t,j} }}\bigg/{\left(\alpha T + \sum\limits_{t = 1}^{T - 1} {z_{t,i}^2}\right) }, \quad W^{ZZ}_{T,ii} = 0.
\end{align}

Iteration \eqref{HTdynamics} can be implemented by neuronal dynamics in a single-layer two-population network, Figure 1E. In this interpretation, ${\bf y}_{T}$ represents the activities of output neurons, which we identify with principal neurons in neuroscience terminology. Similarly, ${\bf z}_{T}$ represents the activities of neurons which connect only within the layer, which we identify with interneurons in neuroscience terminology. Again, although the optimization problems \eqref{HT} and \eqref{HTOnline} did not contain synaptic weights explicitly, we recovered expressions ${\bf W}^{YX}_T$, ${\bf W}^{YY}_T$, ${\bf W}^{ZY}_T$, ${\bf W}^{YZ}_T$ and  ${\bf W}^{ZZ}_T$ that are naturally identified as the weights of synaptic connections in the network.

Finally, by rewriting \eqref{HTWM} in a recursive form, we obtain a fully online algorithm:
\begin{align}\label{HTupdate}
D^Y_{T+1,i} &\leftarrow D^Y_{T,i}+ \alpha, \qquad 
D^Z_{T+1,i} \leftarrow D^Z_{T,i}+ \alpha + z_{T,i}^2\nonumber\\
{W^{YX}_{T+1,ij}} &\leftarrow {W^{YX}_{T ,ij}} + \left( y_{T,i}x_{T,j} -\alpha {W^{YX}_{T,ij}} \right)/D^Y_{T+1,i} \nonumber\\
{W^{YZ}_{T+1,ij}} &\leftarrow {W^{YZ}_{T ,ij}} + \left( y_{T,i}z_{T,j} - \alpha {W^{YZ}_{T,ij}} \right)/D^Y_{T+1,i} \nonumber\\
{W^{YY}_{T+1,ij\neq i}} &\leftarrow {W^{YY}_{T ,ij}} + \left( \gamma y_{T,i}y_{T,j} - \alpha {W^{YY}_{T,ij}} \right)/D^Y_{T+1,i} \nonumber\\
{W^{ZY}_{T+1,ij}} &\leftarrow {W^{ZY}_{T,ij}} \nonumber \\ & \quad +  \left( z_{T,i}y_{T,j} - \left(\alpha + z_{T,i}^2\right) {W^{ZY}_{T,ij}} \right)/D^Z_{T+1,i}\nonumber \\
{W^{ZZ}_{T+1,i,j \ne i}} &\leftarrow {W^{ZZ}_{T,ij}} \nonumber \\ & \quad+ \left( z_{T,i}z_{T,j} - \left(\alpha+ z_{T,i}^2\right) {W^{ZZ}_{T,ij}} \right)/D^Z_{T+1,i}.
\end{align}

To summarize, equations \eqref{HTdynamics} and \eqref{HTupdate} define a neural network algorithm that solves the optimization problem \eqref{HTOnline} for streaming data by alternating between two phases: neural activity dynamics and synaptic updates. After a data sample is presented at time $T$, the algorithm goes into the neuronal activity phase \eqref{HTdynamics}, where neuron activities are updated until convergence to a fixed point. Then, in the second phase of the algorithm, synaptic weights are updated, according to local Hebbian rules \eqref{HTupdate} for  ${\bf W}^{YX}_T$ and  ${\bf W}^{ZY}_T$ connections, and according to local anti-Hebbian rules for  ${\bf W}^{YY}_T$, ${\bf W}^{YZ}_T$ and  ${\bf W}^{ZZ}_T$ connections. 

A similar network was derived in \cite{pehlevan2015NIPS} from the objective \eqref{HTOnline} without the decorrelating term, i.e. $\gamma = 0$. The addition of the decorrelating term does not spoil the locality of learning rules, however, it changes the network architecture by adding anti-Hebbian lateral connections between principal neurons. These new lateral synapses decorrelate neuronal output, whereas in \cite{pehlevan2015NIPS} the output was in general correlated.

In our discussion of the solutions to the offline objective \eqref{HTreg}, we observed that when the number of output channels is larger than the number of output eigenvalues, decorrelation forces extra channels to be silent. In the online case, synaptic weights to silent neurons will eventually decay to zero, as can be seen from inspecting \eqref{HTupdate}.

\subsection{Online whitening}

Finally, we consider the following minimax problem in the online setting:
\begin{align}\label{whitenOnline}
&\{{\bf y}_T,{\bf z}_T\} \leftarrow \mathop {\arg \min }\limits_{{\bf y}_T}  \mathop {\arg \max }\limits_{{\bf z}_T}\,\rm{Tr}\left(- {\bf X}^\top{\bf X}{\bf Y}^\top{\bf Y}  + \alpha T{\bf Y}^\top{\bf Y}\right. \nonumber \\
& \left.+ {\bf Y}^\top{\bf Y}{\bf Z}^\top{\bf Z} - \beta T{\bf Z}^\top{\bf Z} \right)+ \frac \gamma 2 \left\Vert {\rm off}\left({\bf Y}{\bf Y}^\top \right) \right\Vert_F^2 .
 \end{align}
By keeping only those terms that depend on ${\bf y}_T$ or ${\bf z}_T$ and taking the large-$T$ limit, we get:
\begin{align}\label{whitenOnlineReduced}
\{{\bf y}_T,{\bf z}_T\}\leftarrow &\mathop {\arg \min }\limits_{{\bf y}_T}  \mathop {\arg \max }\limits_{{\bf z}_T} \left[- 2{{\bf x}^\top_T}\left( {\sum\limits_{t = 1}^{T - 1} {{\bf x}_t{{\bf y}^\top_t}} } \right){\bf y}_T  \right. \nonumber \\ 
&\left. +{{\bf y}^\top_T}\left( \gamma\,{\rm off}\left({\sum\limits_{t = 1}^{T - 1} {{\bf y}_t{{\bf y}^\top_t}} }\right) +\alpha T {\bf I}_k \right){\bf y}_T \right. \nonumber \\ &  + 2{{\bf y}^\top_T}\left( {\sum\limits_{t = 1}^{T - 1} {{\bf y}_t{{\bf z}^\top_t}} } \right){\bf z}_T -\beta T{{\bf z}^\top_T}{\bf z}_T\bigg].
 \end{align}
As before, this objective is convex in ${\bf y}_T$ and concave in ${\bf z}_T$. The solution of this minimax problem is a saddle-point of the objective function:
\begin{align}
\left( \gamma\,{\rm off}\left({\sum\limits_{t = 1}^{T - 1} {{\bf y}_t{{\bf y}^\top_t}} } \right)+\alpha T{\bf I}_k\right){\bf y}_T &= \left( {\sum\limits_{t = 1}^{T - 1} {{\bf y}_t{{\bf x}^\top_t}} } \right){\bf x}_T \nonumber \\ &\qquad- \left( {\sum\limits_{t = 1}^{T - 1} {{\bf y}_t{{\bf z}^\top_t}} } \right){\bf z}_T, \nonumber \\
\beta T{\bf z}_T& = \left( {\sum\limits_{t = 1}^{T - 1} {{\bf z}_t{{\bf y}^\top_t}} } \right){\bf y}_T.
\end{align}
To obtain a neurally plausible algorithm, we solve these equations by a weighted Jacobi iteration:
\begin{align}\label{whitendynamics}
{\bf y}_{T} &\leftarrow  \left(1-\eta \right){\bf y}_{T} + \eta \left({\bf W}^{YX}_{T}{\bf x}_{T}  - {\bf W}^{YZ}_{T} {\bf z}_{T}- {\bf W}^{YY}_{T} {\bf y}_{T}\right), \nonumber \\
 {\bf z}_{T} &\leftarrow  \left(1-\eta \right){\bf z}_{T} + \eta {\bf W}^{ZY}_{T}{\bf y}_{T},
\end{align}
where,
\begin{align}\label{whitenWM}
W^{YX}_{T,ij} &= \frac 1{\alpha T}{{\sum\limits_{t = 1}^{T - 1} {y_{t,i}^{}x_{t,j}^{}} }},  \quad {W}^{YZ}_{T,ij} = \frac 1{\alpha T}{{\sum\limits_{t = 1}^{T - 1} {y_{t,i}^{}z_{t,j}^{}} }}, \nonumber \\ W^{YY}_{T,i,j\neq i}&= \frac \gamma{\alpha T}{ {\sum\limits_{t = 1}^{T - 1} {y_{t,i}y_{t,j}} }}, \qquad {W}^{YY}_{T,ii} = 0,\nonumber \\
W^{ZY}_{T,ij} &= \frac 1{\beta T}{{\sum\limits_{t = 1}^{T - 1} z_{t,i}y_{t,j} }}.
\end{align}

Again, iteration \eqref{whitendynamics} can be implemented by neuronal dynamics in a single-layer two-population network, Figure 1F, where ${\bf y}_{T}$ represents the activity of principal neurons and ${\bf z}_{T}$ represents the activities of interneurons. Once again, although the optimization problems \eqref{whiten} and \eqref{whitenOnline} did not contain synaptic weights explicitly,  we recovered expressions ${\bf W}^{YX}_T$, ${\bf W}^{YY}_T$, ${\bf W}^{ZY}_T$ and ${\bf W}^{YZ}_T$ which are naturally identified as the weights of synaptic connections in the network. Note that, unlike in \eqref{HTdynamics}, interneurons do not synapse with each other. 

Finally, by rewriting \eqref{whitenWM} in a recursive form, we obtain a fully online algorithm:
\begin{align}\label{whitenUpdate}
D^Y_{T+1,i} &\leftarrow D^Y_{T,i}+ \alpha, \qquad 
D^Z_{T+1,i} \leftarrow D^Z_{T,i}+ \beta \nonumber\\
{W^{YX}_{T+1,ij}} &\leftarrow {W^{YX}_{T ,ij}} + \left( y_{T,i}x_{T,j} - \alpha {W^{YX}_{T,ij}} \right)/D^Y_{T+1,i} \nonumber\\
{W^{YZ}_{T+1,ij}} &\leftarrow {W^{YZ}_{T ,ij}} + \left( y_{T,i}z_{T,j} - \alpha {W^{YZ}_{T,ij}} \right)/D^Y_{T+1,i} \nonumber\\
{W^{YY}_{T+1,i,j\neq i}} &\leftarrow {W^{YY}_{T ,ij}} + \left( \gamma y_{T,i}y_{T,j} - \alpha {W^{YY}_{T,ij}} \right)/D^Y_{T+1,i} \nonumber\\
{W^{ZY}_{T+1,ij}} &\leftarrow {W^{ZY}_{T,ij}} +  \left( z_{T,i}y_{T,j} - \beta W^{ZY}_{T,ij} \right)/D^Z_{T+1,i}.
\end{align}

To summarize, equations \eqref{whitendynamics} and \eqref{whitenUpdate} define a neural network algorithm that solves the optimization problem \eqref{whitenOnline} for streaming data by alternating between two phases: neural activity dynamics and synaptic updates. After a data sample is presented at time $T$, the algorithm goes into the neuronal activity phase \eqref{whitendynamics}, where neuron activities are updated until convergence to a fixed point. Then, in the second phase of the algorithm, synaptic weights are updated, according to local Hebbian rules \eqref{whitenUpdate} for  ${\bf W}^{YX}_T$ and  ${\bf W}^{ZY}_T$ connections, and according to local anti-Hebbian rules for  ${\bf W}^{YY}_T$ and  ${\bf W}^{YZ}_T$ connections. 

A similar network was derived in \cite{pehlevan2015NIPS} from the cost \eqref{whitenOnline} without the decorrelating term, i.e. $\gamma = 0$. The addition of the decorrelating term does not spoil the locality of learning rules, however, it changes the network architecture by adding anti-Hebbian lateral connections between principal neurons. These new lateral synapses decorrelate neuronal output, whereas in \cite{pehlevan2015NIPS} output was decorrelated only if the output was full rank, i.e. dimensionality of principal neural activity was the same as the number of principal neurons.

In our discussion of the solution to the offline whitening objective, \eqref{whitenreg}, we observed that when the number of output channels is larger than the number of output eigenvalues, decorrelation forces extra channels to be silent. In the online case, synaptic weights to silent neurons will eventually decay to zero, as can be seen by inspecting \eqref{whitenUpdate}.

\section{Numerical experiments}\label{sec4}

In this section, we evaluate the performance of the proposed algorithms on a synthetic dataset, which is generated by an $n=64$ dimensional colored Gaussian distribution with a specified covariance matrix. The top 4 eigenvalues are $\lambda_{1..4}=\lbrace 7,6,5,4\rbrace$ and the remaining $\lambda_{5..60}$ are sampled uniformly from the interval $[0,0.5]$. Correlations are introduced in the covariance matrix by generating random orthonormal eigenvectors. For all three algorithms, we choose $\alpha =1$, $\gamma = \lbrace 0,0.5,1\rbrace$, and, for the whitening algorithm, we choose $\beta=2$. In the $T\to\infty$ limit, the optimal non-zero offline eigenvalues  are $\lbrace 7,6,5,4\rbrace$ for PCA and  $\lbrace 2,2,2,2\rbrace$ for whitening. In all simulated networks, the number of principal neurons, $k=10$, and, for adaptive PCA and whitening algorithms, the number of interneurons, $l=10$.  Synaptic weight matrices were initialized randomly, and synaptic update learning rates, $1/D^Y_{0,i}$ and $1/D^Z_{0,i}$ were initialized to 0.01. Network dynamics is run with a weight $\eta=0.1$ until the relative change in ${\bf y}_T$ and ${\bf z}_T$ in one cycle is $<10^{-5}$. 

\begin{figure}
\vskip5pt
\centering
\includegraphics[scale = 0.89]{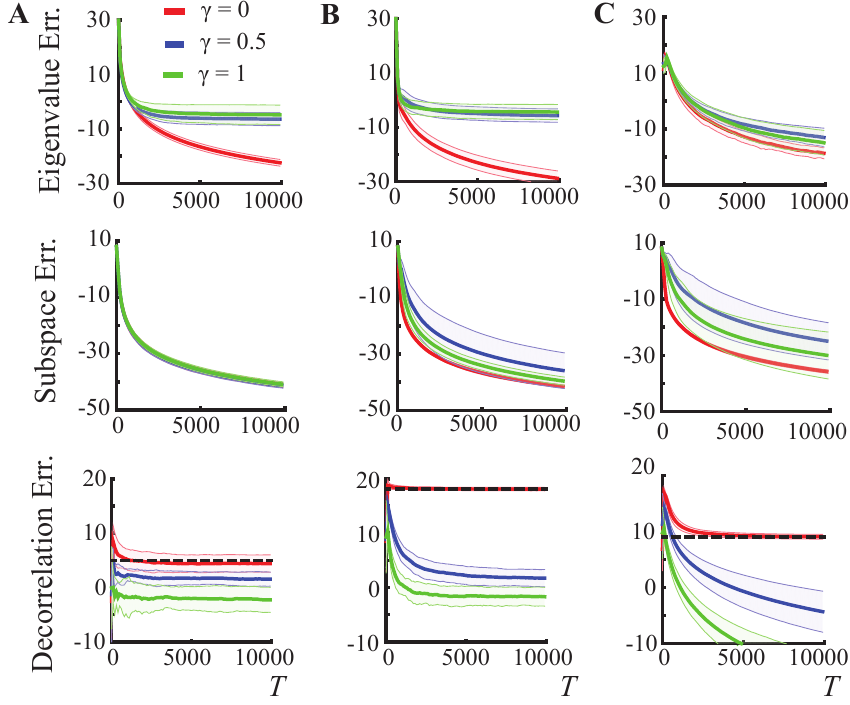}
 \caption{\footnotesize \label{Fig2}Performance of the three similarity matching neural networks - PCA ({\bf A}), adaptive PCA ({\bf B}), and whitening ({\bf C}) - as a function of the number of synthetic data sample presentations (see text). Top: eigenvalue error; middle: subspace error; bottom: decorrelation error (for definitions see text). Means (solid lines) and STDs over 20 runs (shades) of the metrics are shown for three different decorrelation parameters: $\gamma = 0$, or no decorrelation (red), $\gamma = 0.5$ (green), $\gamma = 1$ (blue). Horizontal dashed black lines (bottom row) show the correlation error for random covariance matrices (see text).}
\end{figure}

We characterize the performance of our algorithms using three different metrics. The first metric, eigenvalue error, measures the deviation of the eignevalues of the output covariance $\frac{1}{T} {\bf Y}{\bf Y}^\top$ at time $T$ from their optimal offline values, $10\log_{10}\sum_{i=1}^T(\bar\lambda^Y_{T,i}-\bar\lambda^Y_{{\rm offline},i})^2$ dB. Here $\bar\lambda^Y_{T,i}$ is the $i^{\rm th}$ eigenvalue of $\frac{1}{T} {\bf Y}{\bf Y}^\top$ and $\bar\lambda^Y_{{\rm offline},i}$ is its optimal value. For all three algorithms, the eigenvalue error decreases with time, Figure \ref{Fig2}. Note, however, that adding the decorrelating term, i.e. increasing $\gamma$ leads to a slower decrease of the eigenvalue error.

The second metric, subspace error, quantifies the deviation of the learned subspace from the true principal subspace. To form such metric, at each $T$, we calculate the linear transformation that maps inputs, ${\bf x}_T$, to outputs, ${\bf y}_T = {\bf F}_T{\bf x}_T$ at the fixed points of the neural dynamics stages  of the three algorithms\eqref{dynamics}, \eqref{HTdynamics}, \eqref{whitendynamics}. For PCA ${\bf F}_{T} = \left({\bf I}_k + {\bf W}^{YY}_{T} \right)^{-1}{\bf W}^{YX}_{T}$, for adaptive PCA ${\bf F}_{T} = \left({\bf I}_k+{\bf W}^{YY}_T+{\bf W}^{YZ}_T\left({\bf I}_l + {\bf W}^{ZZ}_{T} \right)^{-1}{\bf W}^{ZY}_T \right)^{-1}{\bf W}^{YX}_{T}$, and for whitening ${\bf F}_{T} = \left({\bf I}_k+ {\bf W}^{YY}_{T} + {\bf W}^{YZ}_T{\bf W}^{ZY}_T \right)^{-1}{\bf W}^{YX}_{T}$. Then, at each $T$, the subspace error is $10\log_{10}\left\Vert{\bf F}_{4,T}{\bf F}_{4,T}^\top-{\bf V}^X_{4,T}{{\bf V}^X_{4,T}}^\top\right\Vert_F^2$ dB, where ${\bf F}_{4,T}$ is an $n\times 4$ matrix whose columns are the top $4$ right singular vectors of ${\bf F}_T$, ${\bf F}_{4,T}{\bf F}_{4,T}^\top$ is the projection matrix to the subspace spanned by these singular vectors, ${\bf V}^X_{4,T}$ is an $n\times 4$ matrix whose columns are the principal eigenvectors of the input covariance matrix ${\bf C}$, ${{\bf V}^X_{4,T}}{{\bf V}^X_{4,T}}^\top$ is the projection matrix to the principal subspace. Figure \ref{Fig2} shows that subspace error decreases quickly with time for all algorithms, however, increasing $\gamma$ leads to a loss of performance for the adaptive PCA and whitening algorithms.

The third metric, decorrelation error, represents correlations among output channels:  $10\log_{10} \left\Vert \frac 1 T {\rm off}\left({\bf Y}_T {\bf Y}_T^\top\right) \right\Vert_F^2$dB, Figure \ref{Fig2}. For $\gamma>0$ output channels decorrelate with the rate increasing with $\gamma$.

For  $\gamma=0$,  the observed output correlation approaches that for a random projection onto the principal subspace (horizontal dashed black lines in Figure \ref{Fig2}). The decorrelation errors for random projections are averaged over a set of 100000 randomly generated $k \times k$ covariance matrices. Each instance of such covariance matrix is constructed from the eigenvalue decomposition, $ {\bf U} {\bf \Lambda} {\bf U}^\top$, where diagonals of ${\bf \Lambda}$ contain optimal offline eigenvalues in $T\to\infty$ limit, and $k\times k$ orthogonal matrices ${\bf U}$ are randomly sampled under the Haar measure.

\section{Dropout of underutilized neurons}\label{sec5}

\begin{figure}
\centering
\includegraphics[scale = 0.80]{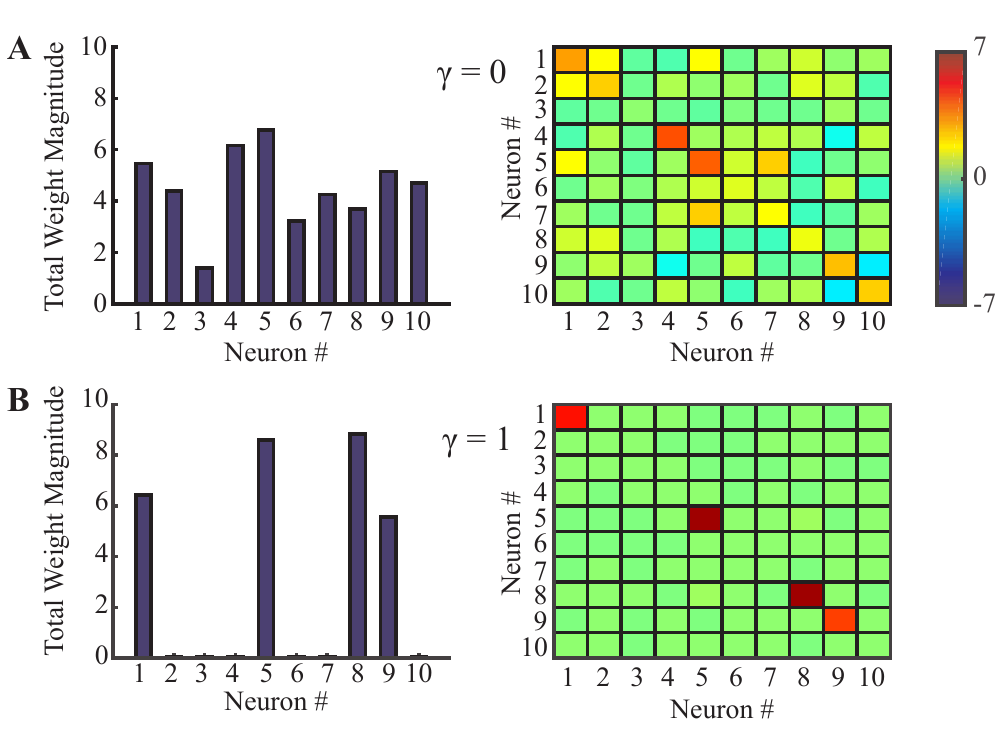}
 \caption{\footnotesize \label{Fig3}Dropout of underutilized neurons in the adaptive PCA network. Simulations of the adaptive PCA network without the decorrelation term, $\gamma =0$, demonstrating that all neurons are active and correlated ({\bf A}) and with the decorrelation term, $\gamma =1$, demonstrating the silencing of extra neurons whose synaptic weights decay to zero ({\bf B}). Left: Summed squared weights of synapses onto principal neurons at $T=10000$, defined for the $i^{\rm th}$ neuron as $\sqrt{\sum_{j=1}^n({W^{YX}_{ij}})^2 + \sum_{j=1}^l({W^{YZ}_{ij}})^2+\sum_{j=1, j\neq i}^k({W^{YY}_{ij}})^2}$. Right: Output covariance matrices for principal neurons, $\frac 1 T {\bf Y}_T {\bf Y}_T^\top$,  at $T=10000$. Input data statistics and parameters same as in Section \ref{sec4}.
 }
\end{figure}

A decorrelation of principal neuron activities in adaptive PCA and whitening circuits makes an interesting prediction. If the number of principal neurons is greater than the typical number of output eigenvalues then the extra neurons are typically silent. Because the weight of synapses onto the extra neurons is proportional to their activities \eqref{HTWM}, \eqref{whitenWM} these synapses will be weak or non-existent. This suggests that the extra neurons disconnect or drop out of the circuit and, in a biological system, may be disposed off. An example of this phenomenon for the adaptive PCA network is shown in Figure \ref{Fig3}. Note that our use of the term ``dropout'' is different from random and intermittent silencing of neurons to regularize learning in deep artificial neural networks \cite{srivastava2014dropout}.

A reverse process may also take place. If the number of principal neurons is less than the typical number of eigenvalues exceeding the threshold, in a biological system, extra neurons may be added to the circuit via neurogenesis.

How does the PCA network behave if the input covariance matrix has few non-zero eigenvalues, $m<\min(k,n)$? As above, the $k-m$ principal are silent in the steady state after each data presentation. However, if the weights of synapses onto these principal neurons are initialized randomly, they do not decay to zero according to \eqref{hah}. Therefore, these silent neurons are active during the initial iterations of the dynamics stage \eqref{dynamics}. Furthermore, the learning rates of the silent neurons stay high and they can become active if the input covariance acquires a new non-zero eigenvalue.

\section{Decorrelation of interneurons}
\label{sec6}
Optimal downstream information transmission by principal neurons in adaptive PCA and whitening circuits does not require decorrelation of interneuron activities. Yet, interneuron decorrelation is easily achieved by adding a decorrelating regularizer $- \rho \left\Vert {\rm off}\left({\bf Z} {\bf Z}^\top\right) \right\Vert_F^2$, where $\rho >0$, to adaptive PCA \eqref{HT} and whitening \eqref{whiten} objectives. From the modified objectives one can derive corresponding online algorithms following the derivations presented in Section \ref{sec3}. As before, the steps of these algorithms can be mapped onto the activity of single-layer two-population neural networks with neurally plausible learning rules. In comparison with the corresponding networks presented in Section \ref{sec3}, the modified adaptive PCA network has stronger lateral connections between interneurons and the modified whitening network adds lateral connections between interneurons.

We note that, previously, interneurons have been added to single-layer circuits for dimensionality reduction \cite{plumbley1994subspace} and sparse dictionary learning \cite{king2013inhibitory,zhu2015modeling}. In addition, for sparse dictionary learning, two-layer circuits with local learning rules have been proposed \cite{olshausen1997,koulakov2011,druckmann2012mechanistic}. However, none of these models included interneuron activities as dynamical variables in objective functions as was done here and in \cite{pehlevan2015NIPS}.

\section{Conclusion}

We developed an optimization theory for PCA and whitening neural networks by adding a decorrelating term to the existing objective functions for projecting data onto principal subspace \cite{pehlevan2015MDS,pehlevan2015NIPS} and deriving, from such objective functions, online algorithms that map onto neural networks with local learning rules. Our theory predicts the dropout of underutilized neurons, due to the decay of their synaptic weights.





\section*{ACKNOWLEDGMENT}

We thank Anirvan Sengupta for useful discussions.



\bibliographystyle{IEEEtran}
\bibliography{biblio}

\end{document}